\documentclass[runningheads]{svmult}

\usepackage{physprbb}  

\begin{document}
\title*{Scientific Potential of Enhancing the Integral-Field
Spectrometer SPIFFI with a Large Detector and High Spectral Resolution}
%
%
%
%
\titlerunning{Scientific Potential of a SPIFFI Upgrade}
%
\author{Frank Eisenhauer\inst{1}
\and Paul van der Werf\inst{2}
\and Niranjan Thatte\inst{1}
\and Tim de Zeeuw\inst{2}
\and Matthias Tecza\inst{1}
\and Marijn Franx\inst{2}
\and Christof Iserlohe\inst{1}}
\authorrunning{Frank Eisenhauer et al.}
%
%
\institute{Max-Planck-Institut f\"ur Extraterrestrische Physik,
Giessenbachstrasse, 85748 Garching, Germany \and Leiden Observatory,
P.O. Box 9513, 2300 RA Leiden, The Netherlands}

\maketitle              

\begin{abstract}
SPIFFI is the near-infrared integral-field spectrometer for the
VLT. Assisted by the SINFONI adaptive optics module, the instrument
will be offered to the astronomical community in 2004. We outline the
scientific rationale for infrared integral-field spectroscopy at the
VLT, and specifically for the enhancement of SPIFFI with a larger
detector and higher spectral resolution gratings. We give examples of
a broad variety of astronomical research which will gain specifically
from the high angular and spectral resolution provided by SPIFFI,
including studies of high red-shift galaxies, merging galaxies,
starburst galaxies, superstar clusters, galactic nuclei, extra-solar
planets, and circum-stellar discs.
\end{abstract}

\section{Introduction}

Integral-field spectrometers record simultaneously the spectrum of
every image point of a two-dimensional field of view. The result of an
observation with an integral-field spectrometer is a three-dimensional
data cube, with two spatial dimensions and one spectral dimension. In
this respect data from integral-field spectrometers are similar to the
results from observations with traditional imaging spectroscopy
techniques like slit-scanning with long-slit spectrometers or a
tunable Fabry-Perot-filter in an imaging camera. However,
integral-field spectrometers have two significant advantages over
these classical techniques. First, integral-field spectrometers are
far more efficient in many applications. These are observations, in
which astronomers are interested in the spectrum of a comparatively
small number of spatial pixels of an extended object, but with large
spectral coverage. In such observations long-slit spectra or
Fabry-Perot images waste the majority of pixels on blank sky. Second,
integral-field spectrometers are often easier to calibrate,
specifically in the infrared. The spectra of every image point of the
two-dimensional field of view are recorded simultaneously, and varying
atmospheric transmission and absorption affect all spectra in the same
way. The calibration of adaptive optics observations with classical
spectrometers is even more difficult, because the point spread
function can vary significantly from exposure to exposure.

Because of the obvious advantages of integral-field spectroscopy at
infrared wavelengths, the Max-Planck-Institut f\"ur Extraterrestrische
Physik (MPE) developed in the mid 1990's the worlds first cryogenic
integral field spectrometer 3D (Weitzel et al.\ 1996). This instrument
was successively upgraded to higher spectral resolution, and has been
operated with fast tip-tilt compensation (Thatte et al.\ 1995), and
high-order adaptive optics (Anders et al.\ 1998; Davies et al.\
2000). The success of this instrument led to the development of the
successor SPIFFI (Eisenhauer et al. 2000), foreseen for operation at
the Very Large Telescope (VLT) of the European Southern Observatory
(ESO). The following sections of this article introduce the
instrument, outline the scientific drivers, specifically for an
upgrade to higher spectral resolution and a larger detector, and
sketch the realisation of this upgrade.

\section{Near-Infrared Integral-Field Spectrometer SPIFFI}
 
SPIFFI (SPectrometer for Infrared Faint Field Imaging) is a fully
cryogenic integral-field spectrometer for the near-infrared wavelength
range from 1.1 -- 2.45 $\mu$m. The spectrometer is part of the SINFONI
(SINgle Faint Object Near Infrared Investigation) instrument for the
VLT, which also includes a modified version of the adaptive optics
MACAO (Donaldson et al.\ 2000). SINFONI is a joint project of MPE,
responsible for the design and manufacturing of SPIFFI, and ESO,
responsible for the design and manufacturing of the SINFONI AO
(Adaptive Optics) Module, with a contribution by NOVA to the adaptive
optics module. Here we give a brief summary of the main
characteristics of SPIFFI. An extensive technical description can be
found in Eisenhauer et al.\ (2000), Mengel et al.\ (2000), Tecza et
al.\ (2000a) and references therein.

The heart of SPIFFI is an image slicer (Tecza et al.\ 2000), which
splits the image into 32 small slitlets, and rearranges them into a 30
cm long pseudo slit. This pseudo long slit is then fed into an
infrared spectrometer, which consists of a three-mirror collimator, a
grating wheel with four different diffraction gratings, a lens camera,
and a Rockwell HAWAII 1024$^2$ detector. The gratings are optimized
for the three near-infrared J, H, and K atmospheric bands, and offer a
spectral resolution ranging from approximately 2000--4000. With this
resolution, SPIFFI allows effective avoidance of the atmospheric OH
lines, which dominate the broad-band background at these wavelengths.
Pre-optics provide three different image scales, ranging from 250
mas/pixel for seeing limited observations down to 25 mas/pixel for
adaptive optics assisted observations at the diffraction limit of the
telescope. A so-called sky-spider allows simultaneous observations of
the sky background. All opto-mechanics are cooled with liquid
nitrogen.\looseness=-2

Table 1 summarizes the point-source sensitivity of SPIFFI (Mengel et
al.\ 2000). The limiting magnitudes are calculated for a
signal-to-noise ratio of five at the full spectral resolution of
approximately 4000. For seeing-limited observations, we integrate over
the seeing disc, assuming the median seeing of 0.69 arcsec on Paranal.
The numbers for adaptive optics assisted observations are calculated
for a Strehl ratio of 15\% in J and H Band, and 25\% in K-Band, and
integrating over the diffraction-limited core. We assume a total
integration time of 2 hours (12 exposures of 600s).

\begin{table}[htp]
\caption{Point source sensitivity of SPIFFI.}
\begin{center}
\begin{tabular}{llll}
\hline\noalign{\smallskip}
Wavelength Band & Pixel Scale & Limiting Magnitude &  \\
 & & no OH Avoidance & with OH Avoidance \\
\noalign{\smallskip}
\hline
\noalign{\smallskip}
 J & Seeing & 19.9 & 20.4 \\
   & Adaptive optics & 19.1 & 19.2 \\
 H & Seeing & 18.9 & 20.1 \\
  & Adaptive optics & 18.5 & 18.6 \\
 K & Seeing & 17.9 & 18.0 \\
  & Adaptive optics & 18.0 & 18.0 \\ 
\hline
\end{tabular}
\end{center}
\label{Tabel1}
\end{table}

\section{Scientific Potential of SPIFFI and Drivers for Higher 
Spectral Resolution} 

With the sensitivity of a fully cryogenic instrument, a spectral
resolution of approximately 100 km/s, and the diffraction-limited
angular resolution of an 8~m telescope, SPIFFI and SINFONI will push
forward astronomical research in many areas. A major domain will
certainly be the exploration of galaxy dynamics in the near and far
universe. This is also the area which will gain significantly from a
spectral resolution of approximately 10000, or equivalently 30 km/s.
The following sections summarize the scientific potential of SPIFFI,
with emphasis on the gains provided by higher spectral resolution.

\subsection{High Redshift Galaxies}

For the time being, Lyman-break galaxies are the best studied tracers
of the cosmic star-formation history (Steidel et al.\ 2001). However,
even fundamental properties of these galaxies are still unknown: What
is the spatial extent of the star-forming regions? Are these galaxies
dominated by a rotational supported disc, or by irregular motion? And
what is the dynamical mass to light ratio of these objects? The reason
for our ignorance is the lack of optical emission lines, which are
shifted to the K-band at redshifts larger than three. Only the latest
generation of near-infrared spectrometers at the VLT and Keck can
measure the velocity dispersion and a hint of the rotation curve in a
few of these galaxies (Pettini et al.\ 2001). However, these objects
exhibit an irregular spatial structure, and single slit positions
cannot provide the necessary two-dimensional information for accurate
rotation curve or velocity dispersion measurements. Integral-field
spectroscopy will overcome this restriction. In addition, these
objects often have a size smaller than the slit width in
seeing-limited spectroscopy, and fine pixel scales and higher spatial
resolution as provided by SINFONI and SPIFFI will help in better
understanding the nature of these objects. Since the masses of these
young galaxies are modest, their velocity dispersions are only of
order 50--100 km~s$^{-1}$, and SPIFFI observations of these objects
will require a spectral resolution of approximately 10000.

\subsection{Merging Galaxies}

The evolution of galaxies is strongly affected by interactions with
neighbouring systems. In particular, merging galaxies initiate
dramatic processes, including triggering of starbursts, lighting up of
galactic nuclei, and changes in Hubble-type.  For such complex
systems, imaging spectroscopy is crucial for understanding the
dynamical structure. Because the appearance of merging galaxies is
very much dominated by local dust absorption or star-formation, the
observer may be biased in positioning the slit in traditional
spectroscopy.

NGC 6240 is the local template for a pair of merging galaxies (Tecza
et al.\ 2000, Van der Werf et al.\ 1993). With its total luminosity of
$6\times10^{11}\,L_\odot$, this galaxy nearly fulfils the criteria for
an ultra-luminous infrared galaxy. At a distance of only 97 Mpc, it is
one of the few systems which allows detailed study of the underlying
processes.  With integral-field spectroscopy, the observer can
directly derive radial velocity and velocity dispersion maps for the
entire light distribution. In addition, the variety of spectral
features in the K-band allows the separate investigation of gaseous
(from the $H_2$ emission lines) and stellar components (from the CO
absorption bands) in a homogeneous way. The $H_2$ emission lines have
width of up to $550$ km/s FWHM which must be the superposition of
several narrower lines, as is also the case for the CO absorptions
bands. With higher spectral resolution the individual components will
be resolved. 

\subsection{Starburst Galaxies}

Merging galaxies are extreme examples of how the interaction between
two galaxies impacts their future evolution. But also non-destructive
encounters can alter the appearance of a galaxy in significant
ways. The prototypical starburst galaxy M82 provides an example, in
which the encounter with its neighbor M81 has triggered
extraordinarily strong circumnuclear star-formation. While the global
properties of M82 are well-known from classical infrared and optical
spectroscopy (Rieke et al.\ 1993), the detailed distribution and
history of star-formation cannot be derived from modelling the global
properties alone. Observation of the circum-nuclear starbursts with an
integral-field spectrometer (F\"orster-Schreiber et al.\ 2001) reveals a
complex spatial distribution of star-forming regions with different
ages. This data allows modeling of the individual clusters, which
removes many ambiguities from the global models. The multiplex
advantages of integral-field spectrometers over classical techniques
reduce the necessary observing time significantly for such regions,
because typically a couple of star-forming regions are covered in a
single exposure.

\subsection{Super Star Clusters}
 
The duration of starbursts is typically only a few million years, but
the remnants of the so called super star clusters -- the largest
star-forming regions in starbursts -- may evolve to globular clusters.
A severe counter argument against this scenario, however, might be the
frequency of stellar masses in starbursts and globular clusters.
Although no firm conclusion has been reached for starbursts
(Eisenhauer 2001; Gilmore 2001), there is strong evidence that these
regions form proportionally fewer low mass stars with masses around 1
$M_\odot$ than typically observed in our Galaxy.  Globular clusters,
however, are known to contain a large number of such low-mass
stars. Measurement of the initial mass function in super-star
clusters, or even the ratio of high and low mass stars, would thus
give strong evidence in favour or against the hypothesis of the
evolution of starburst super-star clusters into globular clusters.
While the light traces the high mass stars in these young clusters,
the frequency of low-mass stars can only be measured through the total
mass of the cluster, so that dynamical mass determinations are
required. However, there are only few super star cluster with
dynamical mass measurements, for example in NGC 1569 and NGC 1705 (Ho
\& Filipenko 1996), or in the Antennae galaxy (Mengel et al.\ 2001).
Because many of these clusters are highly obscured, near- infrared
spectroscopy will significantly enlarge the sample. The typical
velocity dispersion of these clusters is several 10 km/s, and they are
marginally resolved in HST images. With a spectral resolution of
approximately 10000, SPIFFI and SINFONI would be ideal for the
observations of these clusters at adaptive optics pixel scales. 

\subsection{Supermassive black holes in Galactic Nuclei}

Dynamical evidence for the presence of large dark mass concentrations
in the nuclei of normal spiral galaxies has been mounting in recent
years (e.g., Gebhardt et al.\ 2000; Ferrarese \& Merritt 2000).  In
most cases, the central dark mass concentrations (presumed to be
supermassive black holes) are inactive or dormant, and their presence
can only be inferred from gas kinematic (e.g., Miyoshi et al.\ 1995;
Marconi et al.\ 2001) or stellar dynamic (e.g., Kormendy et al.\ 1996;
van der Marel et al.\ 1997) measurements.  Gas kinematic measurements
are often hard to interpret due to complex gas motions which can be
perturbed by non-gravitational effects such as shocks, magnetic
fields, inflows etc.  An unambiguous Keplerian velocity profile of the
gas at radii close to the black hole can only be observed in a few
cases, such as the water maser line observations of NGC 4258 by
Miyoshi et al.\ (1995). Stellar dynamics measurements are more robust,
and have been carried out for a number of elliptical galaxies (e.g.,
Kormendy \& Richstone 1995). However, this is difficult to do for
spiral galaxies, due to the high dust extinction blocking direct view
of the nucleus at visible wavelengths. The radius of influence of a
black hole is small (de Zeeuw 2000), so high spatial resolution (or
adaptive optics) is needed to see the increase in central velocity
dispersion indicative of a central dark mass concentration.  The
obscuration in the nuclear regions of most spiral galaxies can prevent
correct identification of the dynamical nucleus, which is often not
the brightest visible (or near infrared) unresolved nuclear source, as
demonstrated for M83 (Thatte, Tecza \& Genzel 2000). In addition, the
stellar orbits are often complex, and two-dimensional velocity
dispersion and rotation maps are required to correctly account for
anisotropy effects. SINFONI can overcome all these limitations and
make a significant impact in establishing the demography of
supermassive nuclear dark masses in nearby spiral
galaxies.\looseness=-2

\subsection{Brown Dwarfs and Extrasolar Planets}

In the next years, all major observatories will have high-order
adaptive optics systems available at their large telescopes, and
certainly many new brown dwarf and giant planet candidates will be
identified in the proximity of nearby stars. The first example of such
an object was the brown dwarf Gliese 229b (Nakajima et al.\ 1995;
Geballe et al.\ 1996). But broad-band photometry alone will not allow
the accurate determination of the spectral type of these
objects. Reliable mass estimates require spectroscopic classification.
However, long-slit spectroscopy of such objects, which are several
magnitudes fainter than the primary components, and which lie in the
remaining seeing halo of the partially-corrected adaptive optics
images, will be difficult, because accurate deconvolution of the
primary and secondary component requires the two-dimensional
information of the underlying point spread function. Integral-field
spectroscopy with SPIFFI and SINFONI will provide the spectra and the
two-dimensional point spread function simultaneously, and will thus
simplify significantly spectroscopy at or close to the diffraction
limit of an 8~m telescope.

\subsection{Circum-stellar Discs}

Planets are supposed to form proto-planetary discs, many of which have
been discovered with the Hubble Space Telescope (McCaughrean 1995).
At the distance of the Orion nebula, the emission from the molecular
hydrogen in such discs typically extends over about one arc-second.
The K-band hydrogen emission lines could be the ideal tracer of the
rotation of the disc, because at these wavelengths the SINFONI
adaptive optics system provides the best correction of the atmospheric
turbulence. However, the present spectral resolution of SPIFFI is too
low for measuring the rotation of these discs. For a 5 solar mass star
in Orion, the Keplerian velocity would be 4 km/s at a radial distance
of 1 arc-second, and 15 km/s at a distance of 0.1 arc-seconds. These
observations would benefit from spectral resolutions (in excess) of
10000.

\section{Advantage of a Larger Detector}

In its present configuration, SPIFFI incorporates a Rockwell HAWAII
1024$^2$ array. In order to make maximum use of the detector and have
a thousand spatial elements, the spectra of SPIFFI are not Nyquist
sampled, but the slit width corresponds to one detector pixel. In
consequence, observations with SPIFFI at its highest spectral
resolution require two exposures, in which the spectra are offset by
half a pixel (Eisenhauer et al.\ 2000). This technique was applied
successfully in the precursor instrument 3D, but requires good
observing conditions and additional data processing. In addition, the
image motion between the two exposures should not exceed a fraction of
a pixel, which puts strong requirements on the tracking accuracy and
mechanical stiffness of the system. An upgrade to the Rockwell HAWAII
2048$^2$ will overcome the problems with the spectral dithering by
providing Nyquist sampled spectra in a single exposure. The number of
spatial elements will not be increased, so that two detector pixels
cover one sky pixel.

An additional advantage of a larger detector is a less stringent
requirement on the image quality of the spectrometer camera. Not only
is the light for a spectral resolution element now spread over two
pixels instead of one pixel, but also the \mbox{f-number} of the
specrometer camera is significantly increased from approximately 1.6
to 3.1. The upgrade to a larger detector would also open an
opportunity for the implementation of an array manufactured using
molecular beam epitaxy.  These arrays are expected to have a
negligible dark current (Kozlowski et al.\ 1998) well below the 0.1
e$^-$/s dark current of the present HAWAII 1024$^2$ arrays, which are
grown in the traditional liquid phase epitaxy technique.  This upgrade
will thus significantly increase the sensitivity in adaptive optics
assisted observations at the smallest pixel scale of 25 mas/pixel, in
which the typical sky brightness in the H-band between the OH lines
corresponds to 0.0015~e$^-$/s.

\section{Upgrade Plan for SPIFFI}

The present planning forsees a standalone guest-phase operation of
SPIFFI without the AO module for 2002, and commissioning of the joint
SPIFFI and SINFONI instrument in late 2003. Regular observations will
start in 2004. Depending on progress of the instrument, we propose the
upgrade of SPIFFI prior to the commissioning of the SINFONI facility
instrument.  

The enhancement of SPIFFI with a larger detector is independent of an
upgrade to higher spectral resolution. A larger spectral resolution
requires the exchange of one or several of the four gratings in
SPIFFI. The design of SPIFFI accounts for this possibility by having a
comparatively large diameter of the collimated beam of approximately
100 mm. In K-band, a spectral resolution of approximately 11000 can be
achieved either with a grating with 100 grooves/mm, operated in fourth
order, or a grating with 200 grooves/mm, operated in second order.
Because of the small separation of adjacent grooves -- twice the
wavelength -- in the latter case, polarization effects become dominant
(Loewen et al.\ 1977) and need to be calculated before a decission on
the grating design can be made. In both cases the larger anamorphic
magnification of such a high resolution grating will restrict
operation to pixel scales smaller than 200 mas, otherwise vignetting
of the spectrometer camera would degrade the sensitivity of the
instrument.

The upgrade to the Rockwell HAWAII 2048$^2$ detector implies exhange
of the spectrometer camera and the detector readout board.  Because
the pixels of the larger detector have almost the same size (18 $\mu$m
vs 18.5 $\mu$m in the HAWAII 1024$^2$ detector), the focal length of
the camera must be increased from approximately 180 mm to 350 mm.
This simplifies the lens optics, but requires a fold mirror because of
the tight design volume available in the SPIFFI cryostat. A
preliminary five-lens design with spherical lenses made from Barium
Fluoride and the Schott glass IRG2 was foreseen since the early design
phase, so that the upgrade to a larger detector is straightforward.

\section{Summary}

The adaptive optics SINFONI and its integral-field spectrometer SPIFFI
will provide unprecedented imaging spectroscopy at the diffraction
limit of an 8~m telescope. The high sensitivity, the broad wavelength
coverage, and several image scales optimize this instrument for the
observation of a variety of objects from the early universe to nearby
exo-planet candidates. However, many applications -- specifically the
observation of the dynamics in complex galaxy systems -- will benefit
strongly from an enhanced spectral resolution of about 10000. In
addition, the upgrade to a larger detector with lower dark current
will simplify observation and data-reduction, and increase the
sensitivity of the instrument at adaptive optics pixel scales.  This
upgrade of SPIFFI is straightforward, and could be implemented in an
early phase of the facility mode operation.

%

\end{document}